\documentclass[conference]{IEEEtran}
\IEEEoverridecommandlockouts
\usepackage{cite}
\usepackage{amsmath,amssymb,amsfonts}
\usepackage{algorithmic}
\usepackage{graphicx}
\usepackage{textcomp}
\usepackage{xcolor}
\usepackage{fancyhdr}
\usepackage{multirow}

\def\BibTeX{{\rm B\kern-.05em{\sc i\kern-.025em b}\kern-.08em
    T\kern-.1667em\lower.7ex\hbox{E}\kern-.125emX}}
\begin{document}

\title{Living-off-the-Land Abuse Detection Using Natural Language Processing and Supervised Learning}

\author{\IEEEauthorblockN{Ryan Stamp}
\IEEEauthorblockA{\textit{School of Computer Science} \\
\textit{University of Guelph}\\
Guelph, ON, Canada \\
rstamp@uoguelph.ca}
}

\maketitle

\begin{abstract}
Living-off-the-Land is an evasion technique used by attackers where native binaries are abused to achieve malicious intent. Since these binaries are often legitimate system files, detecting such abuse is difficult and often missed by modern anti-virus software. This paper proposes a novel abuse detection algorithm using raw command strings. First, natural language processing techniques such as regular expressions and one-hot encoding are utilized for encoding the command strings as numerical token vectors. Next, supervised learning techniques are employed to learn the malicious patterns in the token vectors and ultimately predict the command's label. Finally, the model is evaluated using statistics from the training phase and in a virtual environment to compare its effectiveness at detecting new commands to existing anti-virus products such as Windows Defender.
\end{abstract}


\section{Introduction}
\label{sec:introduction}

As malware detection methods become more sophisticated, adversaries constantly look for new approaches to remain undetected in target networks \cite{a1}. One such method that has become popular in recent years is living-off-the-land (LOL). This technique involves leveraging existing benign binaries (LOLBins) to achieve the desired result. These binaries are typically legitimate system files that are required for normal operating system execution; however, they also possess the ability to perform particular procedures desired by adversaries, such as downloading files from a remote source.

In a study by Barr-Smith et al. \cite{cit:lol_analysis}, a large range of malware samples were analyzed (totaling nearly 32 million samples), and of them, 9.6\% were found to utilize native binaries. This statistic is non-negligible and demonstrates the real threat LOL techniques possess.

However, detecting and/or preventing LOLBin abuse presents unique challenges compared to other malware detection forms. The primary issue is that these binaries are legitimate system files, so existing techniques such as signature detection will not work here. Instead, more sophisticated methods must be used.

In this paper, we propose a novel command string-based detection algorithm. Our hypothesis states that certain tokens, or certain combinations of tokens, will be used more frequently in malicious commands than in benign commands. By employing natural language processing (NLP) techniques such as regular expressions and one-hot encoding, we can parse and represent the tokens of a command string as numerical vectors \cite{a2}. From here, we train a supervised learning model to predict how malicious a token of a command string is likely to be and, ultimately, how malicious the command itself is. Finally, we evaluate the model to determine the effectiveness of the training set and its ability to detect new threats not already detectable by existing anti-virus products.

\section{Background and Related Works}

Nowadays, cybersecurity is a hot topic for many researchers in the academic and industrial sections. We can find a lot of security mechanisms and defenses in attack detection and threat hunting in \cite{a3,b1,a4,a5}. Moreover, a considerable number of studies have proposed other security solutions like machine learning and blockchain to design and implement detection models \cite{a6,a7i}. A few of these security solutions are the detection of outliers and anomalous data and behavior in Internet of Things (IoT) environments, both of which have been of interest to the academic and industrial research communities, respectively \cite{b3,b2}. Studies have looked at the analysis of IoT attacks and the countermeasures based on the devices connected to the intelligent networks\cite{a7}. The detection of outliers in IoTs, Industrial Internet of Things (IIoT), and similar environments is well known for its notable performance using Artificial Intelligence (AI)-based methods, including machine learning-based and deep learning-based approaches.

In order to identify anomalous behavior on IIoT sites, Hasan et al.  \cite{17} have compared the performance of machine learning models by measuring precision, accuracy, f1-score, and recall. For techniques like Artificial Neural Networks (ANN), Random Forest (RF), and Decision Tree (DT), 99.4\% accuracy was reported by the system. Metrics such as f1-score and recall are better performed by the RF model than other techniques. An ensemble technique based on the Average One-Dependence Estimator (AODE) and RF model is presented by Jabbar, and his colleague in  \cite{18}. Based on the Kyoto data, the proposed technique showed a high accuracy of 90.51 percent and a low False Alarm Rate (FAR) of 0.14 percent. Furthermore, AL-Hawawreh et al. \cite{19} have developed a deep learning-based ensemble model for detecting anomalies in IIoT traffic. With the help of TCP/IP data packets, they have used a feed-forward and AE architecture for training and validating.

Research regarding LoLBin detection has mostly been conducted in recent years. The main concern with LoLBin detection is that the adversary may not create any malicious files for anti-virus software to detect \cite{cit:lol_active_learning}\cite{cit:lol_api_signature}; nor can  these binaries simply be blocked since they provide necessary functionally to the operating system. A common example is a certutil.exe binary native to Windows operating systems. This tool is used for managing certificates but possesses the ability to download files that adversaries can exploit. To overcome this issue,  researchers have turned to AI to detect the malicious use of otherwise benign binaries.

In a study by Ongun et al. \cite{cit:lol_active_learning}, the issues with current methods are outlined, relying on command line parsing by regular expressions that yield high false-positive rates. The study proposes an active learning approach where an AI will select samples for an analyst to classify. The AI works under the same hypothesis that we propose; that certain tokens in the command appear more frequently for malicious commands. Although the results from the study are promising, basing decisions on the tokens alone is very binary-specific as different binaries will often use different tokens in their commands to express the same functionality. Our research will improve on this limitation by developing custom parsing rules to determine the functionality of specific commands so that the same patterns can be recognized across different binaries.

We also want to briefly mention that LOLBin detection also has use in the mobile space. In a paper by Bellizzi et al. \cite{cit:lol_android}, Just-in-Time Memory Forensics (JIT-MF) is leveraged to detect and fight against LOLBin abuse in Android systems. However, although mobile forensics and security are arguably just as important as forensics in desktop operating systems, this paper and its proposed method will solely focus on desktop environments.

\section{Methodologies}
\label{sec:methodologies}

This section will discuss the methodologies behind our LOLBin detection. It will discuss the origin of the dataset, the pre-processing and tokenization techniques, the building of the feature vectors and feature matrix, and finally, the classifier model and its prediction interface.

\subsection{Dataset}
\label{sec:dataset}

The dataset consists of raw command strings for a variety of known LOLBins labeled as either benign or malicious. Since there are likely hundreds of LOLBins that malicious parties could abuse, we chose to focus on only a small subset of the most popular. In total, our model supports 16 different LOLBins listed as follows: bitsadmin, certutil, cmstp, csc, cscript, mmc, msiexec, msxsl, reg, regsvcs, regsvr32, rundll32, schtasks, sqlps, wmic, and wscript. Additionally, our methodologies were designed to make the process of adding new LOLBins relatively trivial, and the such process will be discussed in the later sections.

Benign commands for the LOLBins are rather plentiful. Our dataset they were collected from our partner organization's raw dataset of collected process events from numerous client machines. Malicious samples, by comparison, are harder to obtain. The primary struggle is a lack of readily available samples that have been hand-labeled by experienced analysts. However, for our model, we are less interested in learning specific commands and more interested in learning specific patterns.

LOLBAS \cite{cit:lolbas} is an excellent community-run project for documenting the various LOLBins and providing examples of how they can be abused. Since our pre-processing stages will extract the patterns from command strings, we can collect the malicious samples for our dataset from these LOLBAS examples.

The DFIR Report \cite{cit:dfir_report} is another great resource for obtaining malicious commands. This site hosts reports on real intrusions and lists the many commands that were executed. Any commands involving LOLBins from this resource were collected and added to the validation dataset, which is used later during the evaluation of the model.

The final challenge with our dataset is the imbalance of malicious and benign samples. Since the vast majority of commands encountered in the wild will be benign, it follows that our dataset will also possess the same or similar ratio. However, this is an issue as imbalanced datasets can heavily degrade the effectiveness of the model \cite{cit:smote}.

Synthetic Minority Oversampling Technique (SMOTE) \cite{cit:smote} is an algorithm for balancing a dataset. In short, the algorithm generates new feature vectors by creating line segments between minority class neighbors. These line segments allow for the generation of new vectors to balance the dataset but without introducing any new information that could alter the model's training. In practice, this technique proved very effective.
\subsection{Lexers and Tokenization}
\label{sec:lexer}

Since the model operates on raw command strings, custom-built lexers are first employed to tokenize the input strings.

\begin{figure}[htbp]
    \centering
    \includegraphics[width=0.95\linewidth]{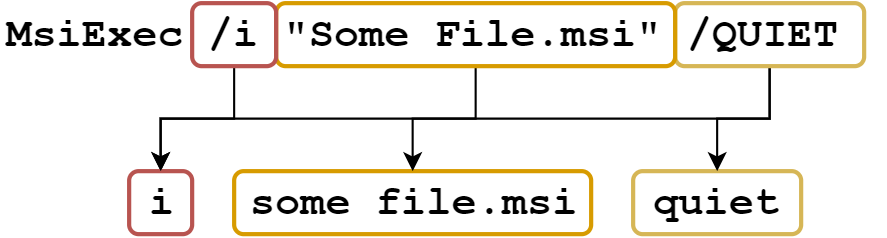}
    \caption{Tokens are normalized}
    \label{fig:normalization}
\end{figure}

The lexers will first split the command string into tokens. Characters separated by whitespace are considered different tokens unless a group of whitespace-separated characters is wrapped in quotations. For example, ``\verb|"Windows Defender"|'' would be considered a single token. Further normalization is applied for each token generated in this manner to decrease the number of unique tokens the model will encounter. This normalization involves removing the leading or trailing quotes, dashes, or slashes. The tokens are also converted to lowercase to enable case insensitive matching later on. Figure \ref{fig:normalization} demonstrates this process visually.

To further decrease the token space, the lexers will next apply several regular expressions to the tokens to detect particular patterns and replace that token with a special token. For example, a url regex is employed to detect if a token is an HTTP or HTTPS url and replaces it with the ``$<$url$>$'' token \cite{a10,a11}. The file regex detects file names and replaces that token with two special tokens, a generic ``$<$file$>$'' token and a specific extension token such as ``$<$ext\_exe$>$''. Figure \ref{fig:lexer} demonstrates this process visually. For a full list of custom tokens, see appendix \ref{apdx:tokens}.

\begin{figure}[htbp]
    \centering
    \includegraphics[width=0.95\linewidth]{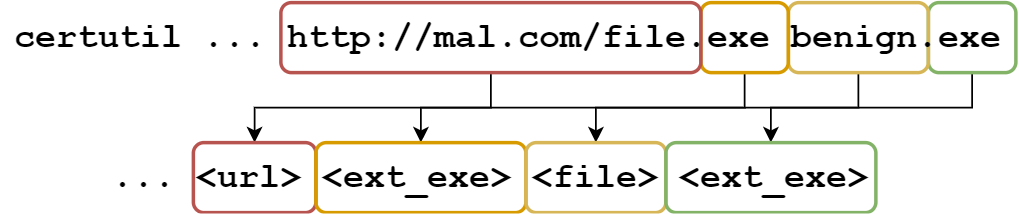}
    \caption{Detected patterns are replaced with custom tokens}
    \label{fig:lexer}
\end{figure}

Do note that since the syntax for every binary can differ greatly, a unique lexer is required for each unique binary. To accomplish this while reusing as much logic as possible, a base class that implements independent binary patterns, such as the URL and file regexes discussed above, was implemented and can simply be instanced to fulfill the requirements for most binaries. However, if necessary, the base class can be extended to more easily add additional pattern recognition if a particular binary requires it. For example, ``rundll32'' implements a subclass to detect javascript parameters.

\subsection{Feature Vectors and Feature Matrix}
\label{sec:feature_vec_mat}

Building feature vectors from command tokens present a unique challenge. Vectors for machine learning models must be numerical \cite{a8,a9}. However, our tokenized strings are words. Thus, some encoding of words to floats must be employed. Although many such techniques exist, the relatively simple one-hot encoding technique was chosen.

Since there are theoretically a nearly infinite number of unique tokens, we must determine which ones are the most important for learning benign versus malicious behavior and only include those tokens in the one-hot encoding. Additionally, we add a special ``$<$rare$>$'' token to act as a dust-bin category for all unrecognized tokens.

To this extent, as with the lexers, we will need a unique feature builder for each unique binary. The reason is that different binaries will have different tokens that are important to determining the command's label and therefore need to be included in the one-hot encoding.

To build the corpus, the entire dataset for the binary is iterated through, and each command is tokenized. Every token that appears in the malicious commands is considered important and is included in the word corpus. Since there are many more unique examples of benign commands compared to malicious samples, only tokens that appear in at least three unique tokenized commands are included in the corpus.

\begin{figure}[htbp]
    \centering
    \includegraphics[width=0.85\linewidth]{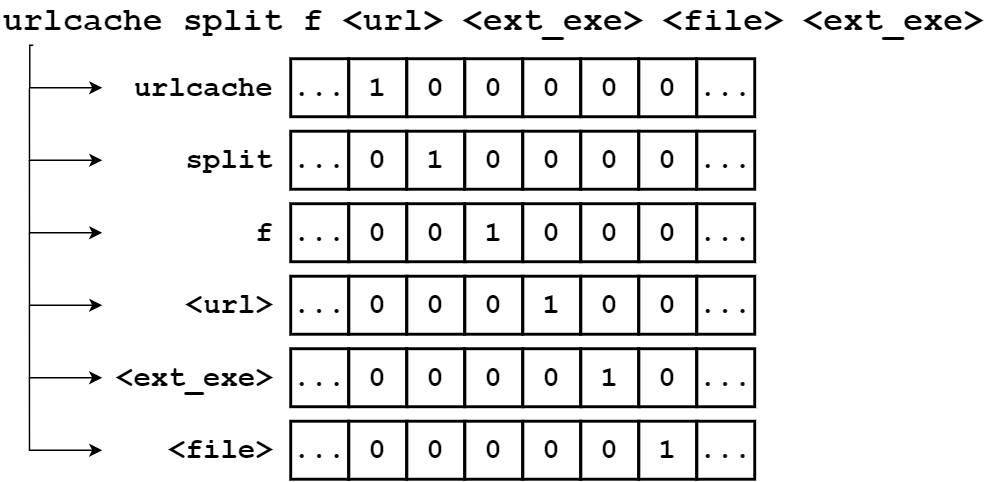}
    \caption{Example one-hot encoding for a collection of tokens}
    \label{fig:one_hot}
\end{figure}

Figure \ref{fig:one_hot} demonstrates visually how a sample tokenized command may be one-hot encoded. Also, since building the word corpus can be expensive depending on the dataset size for a binary, the word corpuses are stored alongside the model artifacts when saving to disk so that they do not need to be regenerated every time.

Now that we can encode our tokens numerically, we next need to build the actual vectors. To achieve better model performance, we not only want to encode what tokens are present in command but also the context of what other tokens appear around them. To achieve this, we employ a novel token feature-building algorithm. Note that this algorithm, as described below, builds a feature vector for each token instead of one vector for the entire command.

\begin{figure}[htbp]
    \centering
    \includegraphics[width=0.85\linewidth]{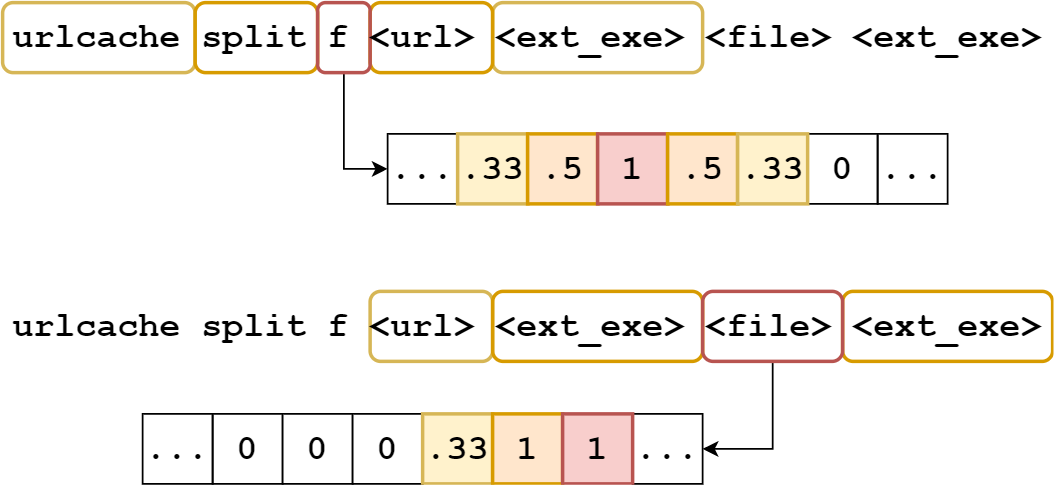}
    \caption{Individual tokens are converted into feature vectors based on what tokens are around them}
    \label{fig:vector}
\end{figure}

Firstly, the feature corresponding to the token being examined is set to one. Next, features corresponding to the tokens away from the target token are set to 1/2. Two away are set assigned values of 1/3 and so on. The window size for how many tokens to consider can be adjusted, but we used a window size of 2. This process is also additive, so if the same token appears more than once in the window around the target token, then this is reflected in the final vector. Figure \ref{fig:vector} demonstrates this algorithm visually.

\subsection{LOLBin Models}
\label{sec:lolbin_model}

So far, we have needed to create unique lexers and feature builders for each unique binary, and the final model is no different. Because the feature vectors are different lengths due to our unique one-hot encoding per binary, a separate model will need to be trained for each binary. At first, this may seem problematic, but this approach boasts several advantages:

\begin{enumerate}
    \item If an existing LOLBin's dataset needs to be updated, then the LOLBin's model can easily be retrained without having to touch the other models.
    \item Similarly, adding support for a new LOLBin can be done simply by adding the new model and the performances of the other models will be completely unaffected.
    \item Re-balancing the dataset using the SMOTE utility (see section \ref{sec:dataset}) can more easily be done on a per binary basis instead of on the entire dataset.
\end{enumerate}

The final training matrices consist of a token vector for every token in every command in the dataset. The train-test split is done with a standard 80\% training and 20\% testing. Note that some more obscure LOLBins have no benign samples in the dataset. In this rare case, that binary model is simply trained on the entire dataset with no splitting or testing.

Two different types of supervised learning models are utilized. SKlearn's multi-layer perceptron (MLP) is the preferred classifier; however, like all neural network-based models, a sufficient amount of data is required to be effective. Suppose the LOLBin in question does not meet the required threshold, in our case, 500 samples. In that case, a Random Forest (RF) classifier is used instead, which can achieve much better performance when less data is available.

\begin{figure}[htbp]
    \centering
    \includegraphics[width=0.95\linewidth]{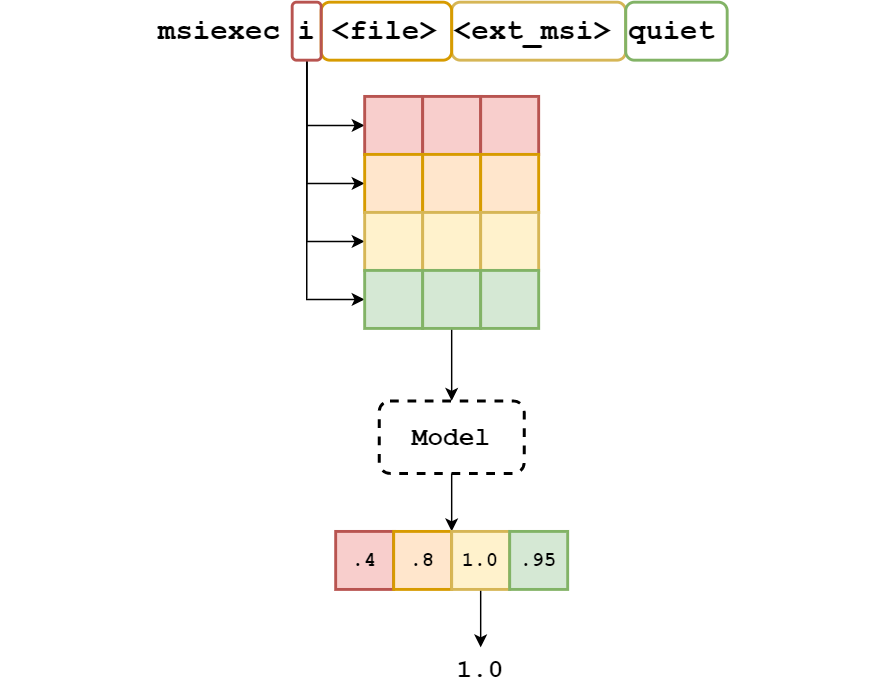}
    \caption{Command score is calculated form multiple token scores}
    \label{fig:matrix}
\end{figure}

Additionally, since the model is trained on token vectors, predictions will be in the form of token scores, not command scores. To obtain command scores, a feature matrix is constructed by combining the token vectors for each token in the command. Submitting the matrix to the model will produce a token score for each row in the matrix. Aggregation techniques can now be employed to interpret the results, such as min, max, and average pooling. In our testing, the max value among the token scores proved to be the most reliable in determining the command's label. Figure \ref{fig:matrix} demonstrates this process visually. Note the token scores used are only examples and may not reflect the true token scores generated by the model.

\subsection{Unimodel}
\label{sec:unimoel}

\begin{figure}[htbp]
    \centering
    \includegraphics[width=0.65\linewidth]{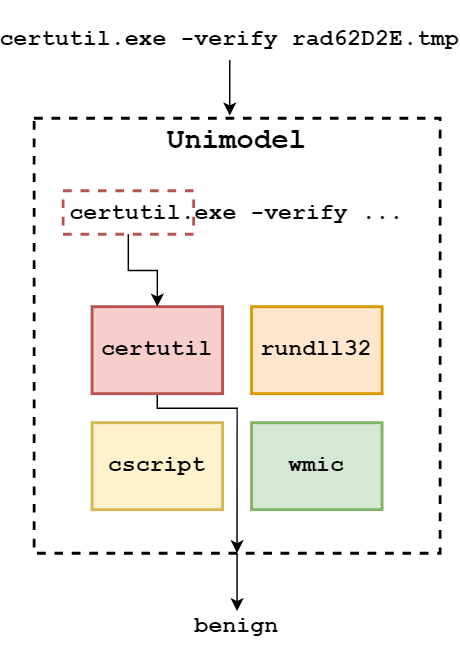}
    \caption{The unimodel encapsulates individual LOLBin models and provides single interface for predictions}
    \label{fig:unimodel}
\end{figure}

Since we trained a separate model for each LOLBin, we have developed a container class, dubbed the "unimodal'', to encapsulate the models into a single object. The unimodeladditionally provides methods for loading the model artifacts from disk and a single interface for making predictions. This method uses techniques such as regular expressions to extract the binary name from the command string and determine which model to query. Figure \ref{fig:unimodel} demonstrates this process visually. The use of the unimodel proved very effective. It solved any issues related to the organization that may arise from using multiple models.

\section{Evaluation}
\label{sec:evaluation}

\subsection{Unimodel Statistics}
\label{sec:model_eval}

The statistics for each LOLBin model are listed in table \ref{tab:model_stats}. Note that the three LOLBins with no benign samples are excluded. It can be seen that the models performed very well overall with near-perfect statistics.

\begin{table}[htbp]
    \caption{Model evaluation in VM results}
    \begin{center}
        \begin{tabular}{|c||c|c|c|c|c|}
            \hline
            & \textbf{Classifier} & \textbf{Accuracy} & \textbf{Precision} & \textbf{Recall} & \textbf{F1 Score} \\ [0.5ex]
            \hline\hline
            \textbf{bitsadmin} & RF & 1.0 & 1.0 & 1.0 & 1.0 \\ [0.5ex]
            \textbf{certutil} & RF & 1.0 & 1.0 & 1.0 & 1.0 \\ [0.5ex]
            \textbf{csc} & MLP & 1.0 & 1.0 & 1.0 & 1.0 \\ [0.5ex]
            \textbf{cscript} & MLP & 1.0 & 1.0 & 1.0 & 1.0 \\ [0.5ex]
            \textbf{mmc} & RF & 1.0 & 1.0 & 1.0 & 1.0 \\ [0.5ex]
            \textbf{msiexec} & MLP & 0.9940 & 0.9884 & 1.0 & 0.9942 \\ [0.5ex]
            \textbf{reg} & MLP & 0.9987 & 1.0 & 0.9973 & 0.9986 \\ [0.5ex]
            \textbf{regsvr32} & MLP & 0.9757 & 1.0 & 0.9451 & 0.9718 \\ [0.5ex]
            \textbf{rundll32} & MLP & 0.9739 & 1.0 & 0.9480 & 0.9733 \\ [0.5ex]
            \textbf{schtasks} & RF & 0.9835 & 0.9836 & 0.9836 & 0.9836 \\ [0.5ex]
            \textbf{sqlps} & RF & 1.0 & 1.0 & 1.0 & 1.0 \\ [0.5ex]
            \textbf{wmic} & MLP & 0.9984 & 0.9968 & 1.0 & 0.9984 \\ [0.5ex]
            \textbf{wscript} & MLP & 1.0 & 1.0 & 1.0 & 1.0 \\ [0.5ex]
            \hline
            \textbf{AVERAGE} & ---  & \textbf{0.9941} & \textbf{0.9976} & \textbf{0.9903} & \textbf{0.9938} \\ [0.5ex]
            \hline
        \end{tabular}
    \end{center}
    \label{tab:model_stats}
\end{table}

These statistics also help to emphasize the advantages of modular models. Since each model is trained specifically for one LOLBin, it can much more easily obtain better accuracy than if it was trained for multiple LOLBins. Separate stats additionally allows analysts to zero in on which models may be underperforming. From here, their datasets can be updated to improve performance; and as mentioned earlier, the models can be trained individually without affecting the others.

\subsection{Virtual Machine Simulation}
\label{sec:vm_sim}

Although statistics can give some idea of how well the model is performing, its real value comes from its ability to detect new commands that are not already detectable by existing solutions \cite{a12}. We also want to validate how effective the model is at detecting new commands not present in the training set, as most of the commands encountered in the wild will be new. For this purpose, we utilize the validation set discussed in section \ref{sec:dataset}.

The model was evaluated against Windows Defender and our partner organization's own Endpoint Detection and Response (EDR) solution. The evaluation consisted of running 83 malicious commands (spanning all LOLBins) in a virtual machine (VM) with both Defender enabled and the EDR agent installed. These 83 commands combine the validation dataset and select commands from the training dataset.

The results are summarized in table \ref{tab:model_eval}. As can be seen, Defender can currently only detect a mere 8 of the malicious LOLBin commands. In contrast, our model can detect all of them. Also, note that for confidential reasons, the EDR results are not disclosed in this paper and are only shared internally with the partner organization.

\begin{table}[htbp]
    \caption{Model evaluation in VM results}
    \begin{center}
        \begin{tabular}{|c||c|c|}
            \hline
            & \textbf{Defender} & \textbf{Model} \\ [0.5ex]
            \hline\hline
            \textbf{bitsadmin} & 0/4 & 4/4 \\ [0.5ex]
            \textbf{certutil} & 2/8 & 8/8 \\ [0.5ex]
            \textbf{cmstp} & 0/2 & 2/2 \\ [0.5ex]
            \textbf{csc} & 0/2 & 2/2 \\ [0.5ex]
            \textbf{cscript} & 0/2 & 2/2 \\ [0.5ex]
            \textbf{mmc} & 0/1 & 1/1 \\ [0.5ex]
            \textbf{msiexec} & 0/5 & 5/5 \\ [0.5ex]
            \textbf{msxsl} & 0/2 & 2/2 \\ [0.5ex]
            \textbf{reg} & 0/19 & 19/19 \\ [0.5ex]
            \textbf{regsvcs} & 1/1 & 1/1 \\ [0.5ex]
            \textbf{regsvr32} & 1/2 & 2/2 \\ [0.5ex]
            \textbf{rundll32} & 2/8 & 8/8 \\ [0.5ex]
            \textbf{schtasks} & 1/10 & 10/10 \\ [0.5ex]
            \textbf{sqlps} & 0/1 & 1/1 \\ [0.5ex]
            \textbf{wmic} & 1/13 & 13/13 \\ [0.5ex]
            \textbf{wscript} & 0/3 & 3/3 \\ [0.5ex]
            \hline
            \textbf{TOTAL} & \textbf{8/83} & \textbf{83/83} \\ [0.5ex]
            \hline
        \end{tabular}
    \end{center}
    \label{tab:model_eval}
\end{table}

This extra level of evaluation provides much greater confidence in the model's effectiveness than utilizing the model statistics alone. Comparing against existing solutions such as Windows Defender and the partner organization's EDR solution additionally helps to prove the model and algorithm's worth and value to detect new threats.

\subsection{False Positives and the Base-Rate Fallacy}
\label{sec:limitations}

In this sub-section, we would like to bring attention to the base-rate fallacy and how it affects the problem of detecting malicious commands.

Axelsson \cite{cit:base_rate_fallacy} excellently describes the base-rate fallacy using the following example. Suppose a disease affects only 1/10000 people in a population, and a test for that disease has a 1\% false-positive rate. Then, if an individual tests positive for the disease using the said test, there is still only a 1/100 chance that they actually have the disease. Intuitively, humans tend to disregard the actual base rate and zero in on the low false-positive rate, so the final results seem rather counter-intuitive. Hence the name ``base-rate \emph{fallacy}''.

We now extend this example to the problem of malicious command detection. Malicious commands make up a tiny percentage of all commands executed on a system, and the rate at which commands are executed is very high. So, by the base-rate fallacy, even if our model (or any other solution) has a very low false positive rate, the resulting hits are still much more likely to be benign than actually malicious.

As part of our evaluation, we ran our model on a live stream of command data generated from numerous clients. The goal was to check for false positives and adjust the models accordingly. However, the sheer amount of false positives was much higher than expected, which we later determined was due to the base-rate fallacy.

This result presents a problem for analysts who monitor alerts generated from such systems. Since the flow of command data to the model is very high, analysts will likely be flooded with false positive alerts that obscure true positives and, in general, consume a lot of resources. Therefore, we recommend that any solution to the malicious command detection problem (whether our presented solution or another) not generate alerts outright, or at the very least, also apply some form of filtering such as whitelisting.

This result also emphasizes the need for human analysts, or more specifically, that this problem may never reach a point of complete automation. Although it may be tempting to adjust the models more to further lower the false positive rate, one would be doing so with the risk of increasing the false negative rate. It is better to generate more false positives and have human analysts available to sift through them than to lower the rate of incorrect alerts and run the risk of missing true malicious activity.

\section{Extensibility}
\label{sec:extensibility}

As stated previously, the unimodel and related procedures were designed to be extendable. This feature is essential in the context of LOLBin detection, as new LOLBins and techniques are being discovered regularly as adversaries constantly adapt their methods to the changing landscape. If adding new patterns and/or retraining the model was a lengthy process, it could act as a bottleneck for the system. Additionally, making sure the unimodel stays modular is important for upholding consistency by not affecting the performance of one LOLBin's detection by adjusting the model of another unrelated LOLBin.
\section{Conclusions and Future Work}
\label{sec:conclusions}

This paper proposed a novel LOLBin command abuse detection algorithm using both regular expressions and supervised learning. Model statistics demonstrate high efficiency at detecting malicious patterns, and further evaluation in a virtual environment shows the model's ability to detect many commands not already detectable by existing anti-virus software such as Windows Defender. The base-rate fallacy was also briefly discussed, and its applications to the LOLBin problem were analyzed.

In terms of future work, support for more LOLBins and/or more token patterns could be added as needed. More sophisticated techniques, in general, may also be required to handle more complex LOLBins such as cmd. Finally, other techniques for reducing false positives could be explored, as was covered in a previous section.

\appendix

\subsection{Custom Tokens}
\label{apdx:tokens}

\textbf{Common (used by all lexers):}
\smallskip

\emph{$<$url$>$} : http or https url \\ eg. \verb|https://malicious[.]com|
\medskip

\emph{$<$url$>$ $<$ext\_*$>$} : url that ends with a path to a file \\ eg. \verb|https://malicious[.]com/file.exe|
\medskip

\emph{$<$ads$>$ }: alternate data stream \\ eg. \verb|benign.txt:malicious.exe|
\medskip

\emph{$<share>$ $<$ext\_*$>$} : shared file \\ eg. \verb|\\10.10.10.10\malicious\file.exe|
\medskip

\emph{$<number>$} : positive or negative integer \\ eg. \verb|100|
\medskip

\emph{$<decimal>$} : positive or negative real number \\ eg. \verb|10.5|
\medskip

\emph{$<ip\_addr>$} : IP address \\ eg. \verb|192.168.1.0|
\medskip

\emph{$<guid>$} : Globally Unique Identifier (GUID) \\ eg. \verb|{e77b42d3-55a5-4b3e-9d08-d59047c2e4c8}|
\medskip

\emph{$<benign\_keyword>$} : LOLBin specific keyword indicating benign behaviour
\medskip

\emph{$<mal\_keyword>$} : LOLBin specific keyword indicating malicious behaviour
\bigskip

\textbf{regsvr32}:
\smallskip

\emph{$<script>$ *} : local or remote script \\ eg. \verb|i:file.sct|
\bigskip

\textbf{rundll32}:
\smallskip

\emph{$<javascript>$} : javascript \\ eg. \verb|javascript:*|
\medskip

\emph{$<dll>$ * *} : DLL file and entrypoint \\ eg. \verb|file.dll,EntryPoint|

\bibliographystyle{ieeetr}
\bibliography{ref}

\end{document}